\documentclass[aps,twocolumn,superscriptaddress,prl,reprint
]{revtex4}

\usepackage{amsmath,amssymb,graphicx}
\usepackage{algorithmic}
\usepackage{enumerate}
\usepackage{color}
\usepackage{times}

\newcommand{\ds}{d_s/2}
\newcommand{\nc}{N_C}

\begin{document}

\title{Chaos in spin glasses revealed through thermal boundary conditions}

\author{Wenlong Wang}
\email{wenlong@physics.umass.edu}
\affiliation{Department of Physics, University of Massachusetts,
Amherst, Massachusetts 01003 USA}

\author{Jonathan Machta}
\email{machta@physics.umass.edu}
\affiliation{Department of Physics, University of Massachusetts,
Amherst, Massachusetts 01003 USA}
\affiliation{Santa Fe Institute, 1399 Hyde Park Road, Santa Fe, New Mexico
87501, USA}

\author{Helmut G. Katzgraber}
\affiliation{Department of Physics and Astronomy, Texas A\&M University,
College Station, Texas 77843-4242, USA}
\affiliation{Materials Science and Engineering Program, Texas A\&M
University, College Station, Texas 77843, USA}
\affiliation{Santa Fe Institute, 1399 Hyde Park Road, Santa Fe, New Mexico
87501, USA}

\begin{abstract}

We study the fragility of spin glasses to small temperature
perturbations numerically using population annealing Monte Carlo. We
apply thermal boundary conditions to a three-dimensional
Edwards-Anderson Ising spin glass.  In thermal boundary conditions all
eight combinations of periodic versus antiperiodic boundary conditions
in the three spatial directions are present, each appearing in the
ensemble with its respective statistical weight determined by its free
energy. We show that temperature chaos is revealed in the statistics of
crossings in the free energy for different boundary conditions.  By
studying the energy difference between boundary conditions at
free-energy crossings, we determine the domain-wall fractal dimension.
Similarly, by studying the number of crossings, we determine the chaos
exponent. Our results also show that computational hardness in spin
glasses and the presence of chaos are closely related.

\end{abstract}

\maketitle

Chaos refers to sensitivity to small perturbations. In addition to
dynamical systems where the phenomenon was first identified, there are
many statistical mechanical systems where chaotic effects have been
predicted and observed. For example,  hysteresis, memory, and
rejuvenation effects found in random elastic manifolds, polymers
\cite{fisher:91d,sales:02,silveira:04,ledoussal:06}, as well as spin
glasses are considered to be a direct manifestation of the presence of
chaos \cite{nordblad:98,dupuis:01,jonsson:04}.  It is surprising and
fascinating  that both the nonequilibrium and equilibrium states of spin
glasses are so fragile to small perturbations.  Chaos is therefore
central to the understanding of both equilibrium and nonequilibrium
properties of spin glasses, as well as related systems. The connection between chaos in spin glasses and dynamical systems has been recently explored~\cite{hu:12}.  Furthermore,
there is mounting evidence that chaos in spin glasses is directly
related to the computational hardness and long thermalization times
\cite{fernandez:13} of these paradigmatic benchmark problems.  As such,
quantifying and understanding chaotic effects in spin-glass-like
Hamiltonians could be of great importance for the development of any
novel algorithm or computing architecture
\cite{zhu:15a,katzgraber:15,hen:15}.

In this work we study the effects of  thermal perturbations. {\em
Temperature chaos}  refers to the property that a small change in
temperature results in a complete reorganization of the equilibrium
configuration of the system.  Temperature chaos has long been predicted
for spin glasses \cite{mckay:82,parisi:84,fisher:86,bray:87}. Although
some early studies raised doubts about the existence of temperature
chaos \cite{billoire:00}, increasing numerical evidence for temperature
chaos has emerged in recent years for various models such as the
random-energy random-entropy model \cite{krzakala:02} and also  more
realistic three- and four-dimensional Ising spin glasses
\cite{sasaki:05,katzgraber:07,fernandez:13}.  It has been suggested that
temperature chaos would only be observable in spin glasses at very large
system sizes and large changes in the temperature
\cite{aspelmeier:02a,rizzo:03}.  However, some studies
\cite{katzgraber:07} demonstrated the existence of temperature chaos via
scaling arguments.

One direct manifestation of temperature chaos is that the free-energy
difference between two boundary conditions that differ by a domain wall
may change sign as a function of temperature. Previous studies examined
the free-energy difference between periodic and antiperiodic boundary
conditions in a single direction to identify temperature chaos
\cite{thomas:11e,sasaki:05}.  This motivates us to study temperature
chaos using {\em thermal boundary conditions} \cite{wang:14}, in which
all $2^d$ combinations of periodic and anti-periodic boundary conditions
in the $d$ spatial directions appear in a single simulation with their
appropriate statistical weights. Thermal boundary conditions provide a
novel and elegant way to study temperature chaos.

Here we quantitatively investigate temperature chaos using population
annealing Monte Carlo \cite{hukushima:03,machta:10,machta:11,wang:14,wang:15e}.
This simulation approach is ideal to study chaos effects in spin glasses
because multiple boundary conditions can be studied at the same time. We
show that temperature chaos is revealed in the statistics of crossings
in the free energy for pairs of boundary conditions \cite{thomas:11e}
and thus establish both qualitatively and quantitatively the presence of
chaos in spin glasses. Our approach can be applied to a multitude of
problems and, in particular, to the search for hard benchmark instances
for novel computing paradigms \cite{katzgraber:15,hen:15}.

What causes temperature chaos? Temperature chaos results from the
existence of dissimilar classes of configurations with similar free
energies but differing energies and entropies~\cite{fisher:86,bray:87}.
Consider two classes of spin configurations, $\sigma_1$ and $\sigma_2$,
corresponding to distinct basins in the free-energy landscape. Within
each class, all spin configurations are similar but the two classes are
dissimilar and differ by a large relative domain wall.  Let $\Delta
F(T)$ be the free-energy difference at temperature $T$ between these two
classes, with $\Delta F(T)=\Delta E(T)- T\Delta S(T)$ where $\Delta E$
and $\Delta S$ are the energy and entropy, respectively, of the relative
domain wall.  Suppose now that  $\Delta E$ and $\Delta S$ are both much
larger than $\Delta F$ and weakly dependent on temperature; then a
small change in temperature may lead to sign change in $\Delta F$.
Suppose that $\Delta F$, $\Delta E$, and $\Delta S$ all behave as power
laws in the size scale $\ell$ of the relative domain wall separating
spin configurations $\sigma_1$ and $\sigma_2$ with leading behavior
$\Delta F \sim \ell^\theta$  but with $\Delta E \sim \Delta S \sim
\ell^{\ds}$ and $\ds > \theta$.  Here $\theta$ is the stiffness exponent
and $d_s$ is the fractal dimension of the domain wall.  As $\ell$
increases, the temperature perturbation $\delta T$ required to change
the sign of $\Delta F$ decreases, i.e., $\delta T \sim \ell^{-\zeta}$
with the chaos exponent $\zeta$ given by $\zeta= \ds -
\theta$~\cite{bray:87}.

We investigate temperature chaos in the Edwards-Anderson (EA) Ising
spin-glass model \cite{edwards:75}. The EA Hamiltonian is
\begin{equation}
\mathcal{H}=-\sum\limits_{\langle ij \rangle} J_{ij} s_i s_j,
\end{equation}
where ${s_i=\pm1}$ are Ising spins. The sum $\langle ij \rangle$ is over
the nearest-neighbor sites in a cubic lattice with $N = L^3$ sites.
$J_{ij}$ is the interaction between spins $s_i$ and $s_j$, and is chosen
from a Gaussian distribution with mean zero and variance 1. We refer
to each disorder realization as a ``sample.''

We use thermal boundary conditions (TBC) to study temperature chaos in
the EA model. In the TBC ensemble each boundary condition $i$ occurs in
the ensemble with a weight depending on its free energy $F_i$.  The
probability $p_i$ of boundary condition $i$ in the ensemble is given by
$p_i = \exp[-\beta (F_i-F)]$, where $F$ is the total free energy of the
system in TBC and $\beta$ the inverse temperature.  Thermal boundary
conditions were introduced to minimize the finite-size effects due to
domain walls and have proved to be useful in studying the
low-temperature phase of the EA model~\cite{wang:14}.  They have been
used with exact algorithms for finding ground states of two-dimensional
spin glasses \cite{landry:02,thomas:07} (referred to there as
``extended'' boundary conditions). A more restricted version of TBC
using periodic and antiperiodic boundary conditions in only a single
direction was used in
Refs.~\cite{hukushima:99,sasaki:05,sasaki:07b,hasenbusch:93}.

\begin{figure}[htb]
\begin{center}
\includegraphics[scale=0.68]{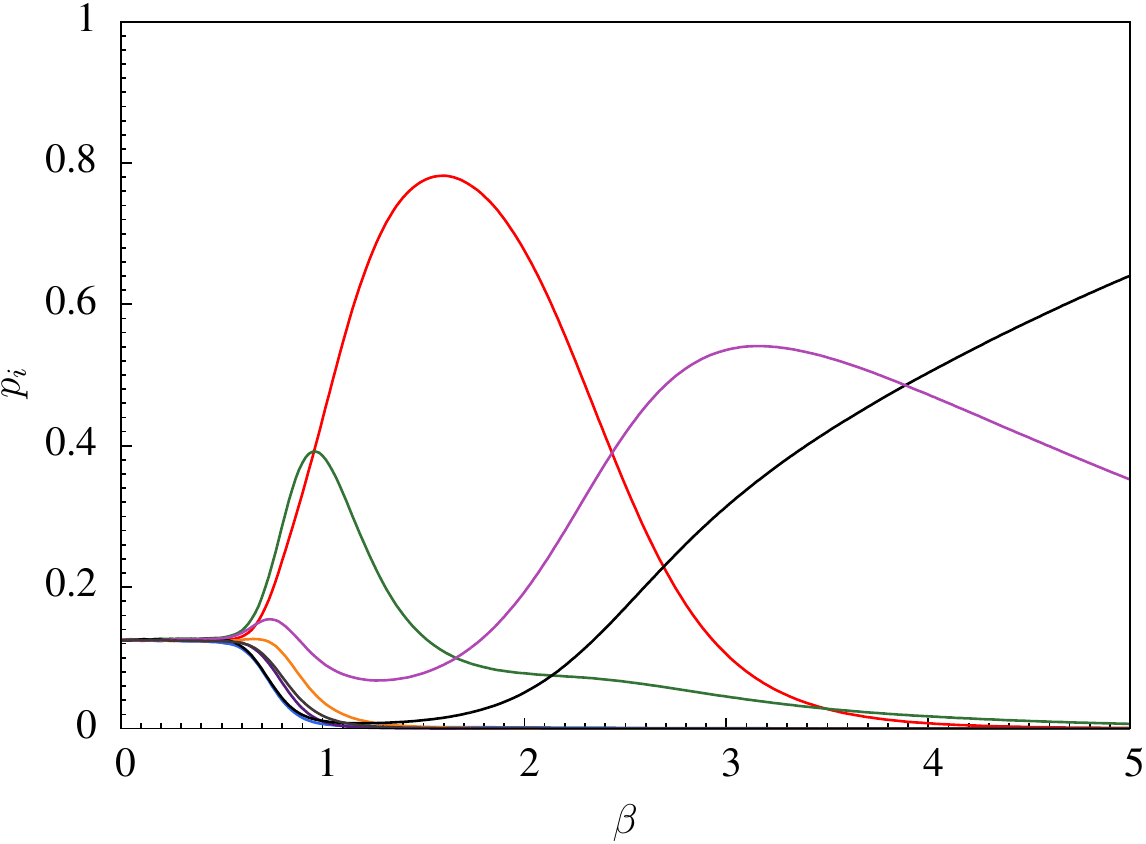}
\caption{(Color online)
A single size $L=10$ sample displaying several boundary-condition
crossing. The plot shows the probabilities of the eight boundary
conditions $\{p_i\}$ as a function of inverse temperature $\beta$. }
\label{fig:cross}
\end{center}
\end{figure}

In thermal boundary conditions, a domain wall on the scale of the linear
system size $L$ separates each boundary condition. Thus temperature
chaos manifests itself as a strong temperature dependence in the
relative free energies of the different boundary conditions (BCs).
Because the stiffness exponent is positive, in the low-temperature phase
one expects that for large systems a single BC will dominate the
ensemble for almost all temperatures. However, as the temperature
changes, the dominant boundary condition will frequently change. A
crossing event occurs when the free-energy difference between two BCs
changes sign. The proliferation of crossing events is a direct
indication of temperature chaos.  Boundary-condition crossing events
between periodic and antiperiodic BCs in one direction were studied in
the two-dimensional EA model in Ref.~\cite{thomas:11e} and identified as
a signature of temperature chaos. Figure \ref{fig:cross} shows BC
probabilities $p_i$ for all eight boundary conditions as a function of
temperature for a single $L=10$ sample. As expected, at high
temperatures, each BC occurs with equal probability. However,  at low
temperatures, four different BCs dominate in different temperature
ranges and, indeed, the dominant boundary condition at the lowest
temperatures has a tiny probability in a range just below the critical
temperature.

We carried out simulations of the three-dimensional EA model in TBC
using population annealing Monte Carlo
\cite{hukushima:03,machta:10,machta:11,wang:14,wang:15e}. Population annealing is
similar to simulated annealing: In both algorithms, the system is cooled
from a high temperature to a low temperature following an annealing
schedule. However, population annealing involves cooling a {\em
population} of replicas and includes a resampling of the population as
it is cooled. At each temperature step in the annealing schedule, each
replica is acted on independently by the Metropolis algorithm. In the
resampling step, which occurs  before the temperature is changed,
replica $i$ is differentially reproduced according to its energy $E_i$.
The expected number of copies of replica $i$ is
$\exp[-(\beta^\prime-\beta) E_i]/Q(\beta,\beta^\prime)$ for a
temperature step from $\beta$ to $\beta^\prime$. The normalization
$Q(\beta,\beta^\prime)$ is chosen such that the expected population size
is unchanged by the resampling step, $Q(\beta,\beta^\prime) = (1/R_0)
\sum_i \exp[-(\beta^\prime-\beta) E_i]$, where $R_0$ is the expected
population size.  The actual number of copies made of replica $i$ is a
random integer whose mean is $\exp[-(\beta^\prime-\beta)
E_i]/Q(\beta,\beta^\prime)$. Note that expected number of copies of each
replica is exactly the reweighting factor between Gibbs distributions at
$\beta$ and $\beta^\prime$. Thus, if the population is representative of
the Gibbs distribution at inverse temperature $\beta$ and $R_0$ is
large, then after resampling, the population is representative the Gibbs
distribution at $\beta^\prime$.  The annealing schedule consists of
$N_T$ temperature steps equally spaced in $\beta$ with $N_S=10$
Metropolis sweeps at each temperature.  Thermal boundary conditions are
easily simulated in population annealing by initializing the population
at $\beta=0$ with $1/8$ of the population in each of the eight BCs~\cite{wang:15e}.
Resampling takes care of making sure that at every temperature, each BC
appears with the correct statistical weight.  We study $2000$ samples of
sizes $L=4$ ($R_0 = 5\,10^4$, $N_T = 101$), $6$ ($R_0 = 2\,10^5$, $N_T =
101$), $8$ ($R_0 = 5\,10^5$, $N_T = 201$), $10$ ($R_0 = 10^6$, $N_T =
301$), and $12$ ($R_0 = 10^6$, $N_T = 301$), down to temperature
$T=0.33$.  The critical temperature is $T_c \approx 0.951$
\cite{katzgraber:06} so the simulations include temperatures that are
deep within the low-temperature phase.  For some hard samples
\cite{wang:14} we use larger population sizes.  In the case of $L=12$
approximately $300$ samples needed to be run with up to a factor of $10$
larger population sizes.

Using population annealing we carry out a quantitative study of boundary
condition crossings.  The temperature difference between crossing scales
as $L^{-\zeta}$ so that the number of crossing $\nc$ in a fixed
temperature interval  scales as $\nc \sim L^\zeta$.  Also, at crossings,
we have that $\Delta F=0$ so that $\ds$ can be obtained from the scaling
of the average of $\Delta E$ at crossings as function of $L$. Finally,
in a previous study we measured the stiffness exponent in TBC.  We
defined the sample stiffness $\lambda$ as $\lambda = \log[f/(1-f)]$
where $f$ is the probability of being in the dominant BC, i.e., $f =
\max_i [ \{p_i \}]$.  We measured $\theta$ as the scaling of the median of
$\lambda$ with system size $L$. Thus, within TBC we can independently
measure all three exponents $\theta$, $\ds$, and $\zeta$ and verify the
relation $\zeta = \ds - \theta$.

Crossings can be divided into two classes: {\em Dominant} crossings are
those such that the two equal BC probabilities at the crossing are
larger than all other BC probabilities. All other crossings are {\em
subdominant}. For large systems, the BC probability at a subdominant
crossing is expected to be typically suppressed by a factor
$\exp(L^{-\theta})$ relative to the dominant BC and thus be increasingly
difficult to observe in TBC simulations. To avoid finite-size
corrections in counting crossings, here we focus on dominant crossings.
On the other hand, for measuring $\Delta E = T\Delta S$ ($\Delta S$ the
change in entropy) at crossings we do not expect a distinction between
dominant and subdominant crossings and, to improve statistics, we use
all crossing with $p_i > 0.05$.

\begin{figure}[htb]
\begin{center}
\includegraphics[scale=0.68]{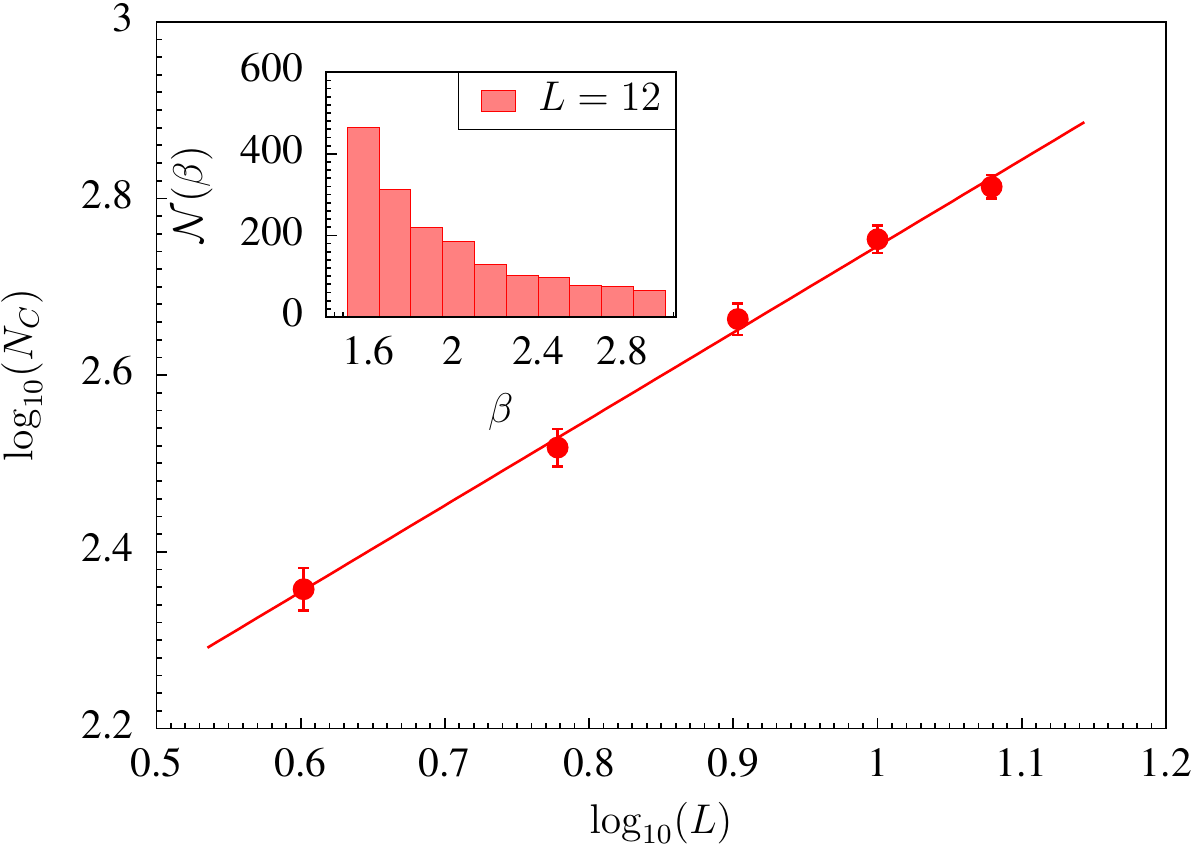}
\caption{(Color online)
Number of dominant crossing in the range $\beta \in (1.5, 3.0)$ vs size
$L$, for $L = 4$, $6$, $8$, $10$, and $12$.  The straight line is the
best power law fit (see text).  The inset is a histogram of the number
of all crossings with $p_i>0.05$ with respect to $\beta$ for system size
$L = 12$.
}
\label{fig:crossvsL}
\end{center}
\end{figure}

Figure \ref{fig:crossvsL} is a log-log plot (base $10$) of $\nc$ vs $L$
where $\nc$ counts  dominant crossing in the range $\beta \in (1.5,
3.0)$.  A simple power-law fit $\nc \sim L^\zeta$ yields $\zeta=
0.96(5)$.  All quoted error bars are one standard deviation statistical errors.
To test the effect of temperature on this exponent, we also
calculated $\zeta$ from two smaller temperature ranges. For $\beta \in
(1.5, 2.0)$ we find $\zeta=1.07(8)$, and from $\beta \in (2.0, 3.0)$ we
find $\zeta=0.85(9)$. For higher temperatures, critical fluctuations may
contaminate the measurement of the chaos exponent while for lower
temperatures the number of crossings is suppressed by the smallness of
the entropy. We note that there is a significant trend to a smaller
value of $\zeta$ at lower temperatures. If one assumes that a single
exponent holds throughout the low-temperature phase, this trend suggests
significant temperature-dependent finite-size corrections.  The inset to
Fig.~\ref{fig:crossvsL} shows a histogram of the number of  crossings in
the range $\beta \in (1.5, 3.0)$ with $p_i > 0.05$ for size $L=12$ as a
function of inverse temperature and reveals that the number of crossings
decreases with temperature, consistent with the fact that the entropy
decreases with temperature so that increasingly large temperature
changes are required to change the free-energy difference between BCs.
In the large-volume limit, the number of dominant crossings per sample
is expected to become infinite but for size $L=12$ temperature chaos
events are infrequent but not rare--there are on average $0.86$
crossings with $p_i>0.05$ per sample in the range $\beta \in (1.5, 3.0)$
and $0.33$ dominant crossings per sample in the same temperature range.
An advantage of using boundary condition crossings in TBC is that
temperature chaos is not a rare event for accessible system sizes in
contrast to overlap correlations in a single boundary condition where
chaotic effects are weak in most samples \cite{fernandez:13}.

\begin{figure}[htb]
\begin{center}
\includegraphics[scale=0.68]{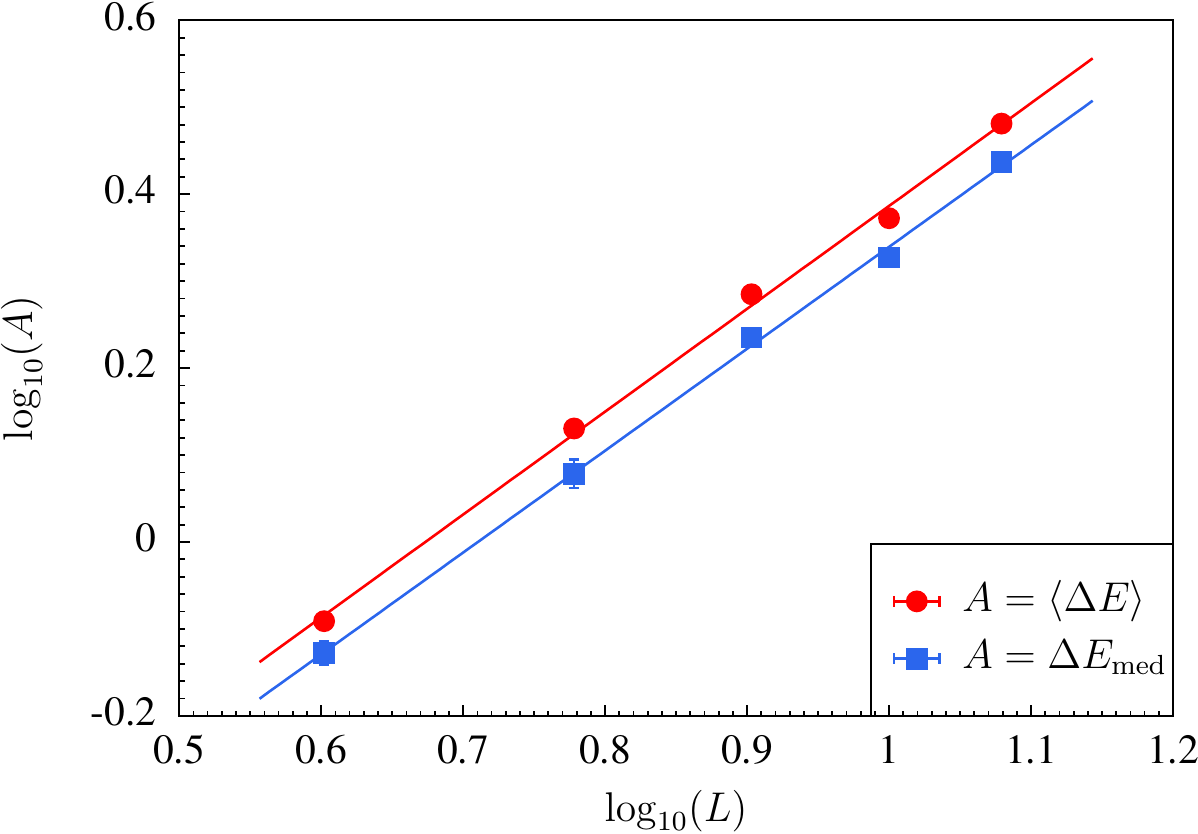}
\caption{(Color online)
Log-log plot of the mean (red circles) and median (blue squares) energy
difference between dominant boundary conditions at crossing in the range
$\beta \in (1.5, 3.0)$ for $L = 4$, $6$, $8$, $10$, and $12$.  The
straight line is the best power law fit (see text).  Error bars are
smaller than the symbols.
}
\label{fig:energy}
\end{center}
\end{figure}

Figure \ref{fig:energy} is a log-log plot (base $10$) of the median and
mean of the absolute energy difference $|\Delta E|$ vs $L$ at all
crossings in the range $\beta \in (1.5, 3.0)$ such that $p_i > 0.05$.  A
simple power-law fit for the mean yields $|\Delta E| \sim L^{\ds}$ with
$\ds = 1.18(2)$ with the same result for the median.  We again test the
effect of the temperature range on $\ds$ by dividing the $\beta$ range
into two intervals, $\beta \in(1.5, 2.0)$ and $\beta \in(2.0, 3.0)$, from
which we obtain the the results for the mean $\ds=1.14(2)$ and
$\ds=1.26(3)$, respectively. There is a significant trend toward larger
values at lower temperatures, suggesting temperature-dependent
finite-size corrections.

Our results for the three-dimensional EA model, $d_s/2=1.17(2)$ and
$\zeta=0.96(5)$, are comparable but slightly smaller than previous work:
For example, $d_s/2=1.29(1)$ was found in Ref.~\cite{palassini:00} based
on perturbations of the ground state, and $d_s/2=1.31(1)$ was found in
Ref.~\cite{katzgraber:01} based on the variance of the link overlap.
Recall that our result for $d_s/2$ in the lower temperature range is
$1.26(3)$, which is within error bars of these zero-temperature results
and suggests large temperature-dependent finite-size corrections.  Our
result, $\zeta=0.96(5)$, is somewhat smaller than $\zeta=1.04$  found in
Ref.~\cite{katzgraber:07} from the spin overlap between different
temperatures. Combined with our estimate of $\theta= 0.27(2)$
\cite{wang:14} we find that the predicted relation $\zeta = \ds -
\theta$ is reasonably well satisfied by our results.

Temperature chaos partially explains why spin-glass simulations are
computationally costly \cite{fernandez:13}. All known efficient
algorithms for equilibrating three-dimensional spin glasses rely on
coupled simulations at many temperatures. Algorithms in this class
include parallel tempering Monte Carlo \cite{hukushima:96}, the
Wang-Landau algorithm \cite{wang:01}, and the algorithm used in this
work, population annealing \cite{hukushima:03,machta:10}.  In these
algorithms, fast mixing at high temperatures provides new configurations
to the low-temperature simulations. Temperature chaos decreases the
effectiveness of these algorithms because the configurations supplied
from higher temperatures are often rather different from the important
configurations at lower temperatures. In TBC, temperature chaos means
that BCs that are important at high temperature are unimportant at low
temperature.   This phenomenon is evident in Fig.~\ref{fig:cross}. One
might worry that boundary conditions that should be important at low
temperature are completely lost at higher temperatures so that the
simulations do not reach the correct TBC equilibrium. To verify that
this is not the case, we performed an additional check of the
equilibration of the TBC ensemble by re-doing several hundred of the
hardest $L=12$ samples using an order of magnitude larger population
sizes in the simulation and we found no difference in the number of
crossings for any sample.

\begin{figure}[htb]
\begin{center}
\includegraphics[scale=0.68]{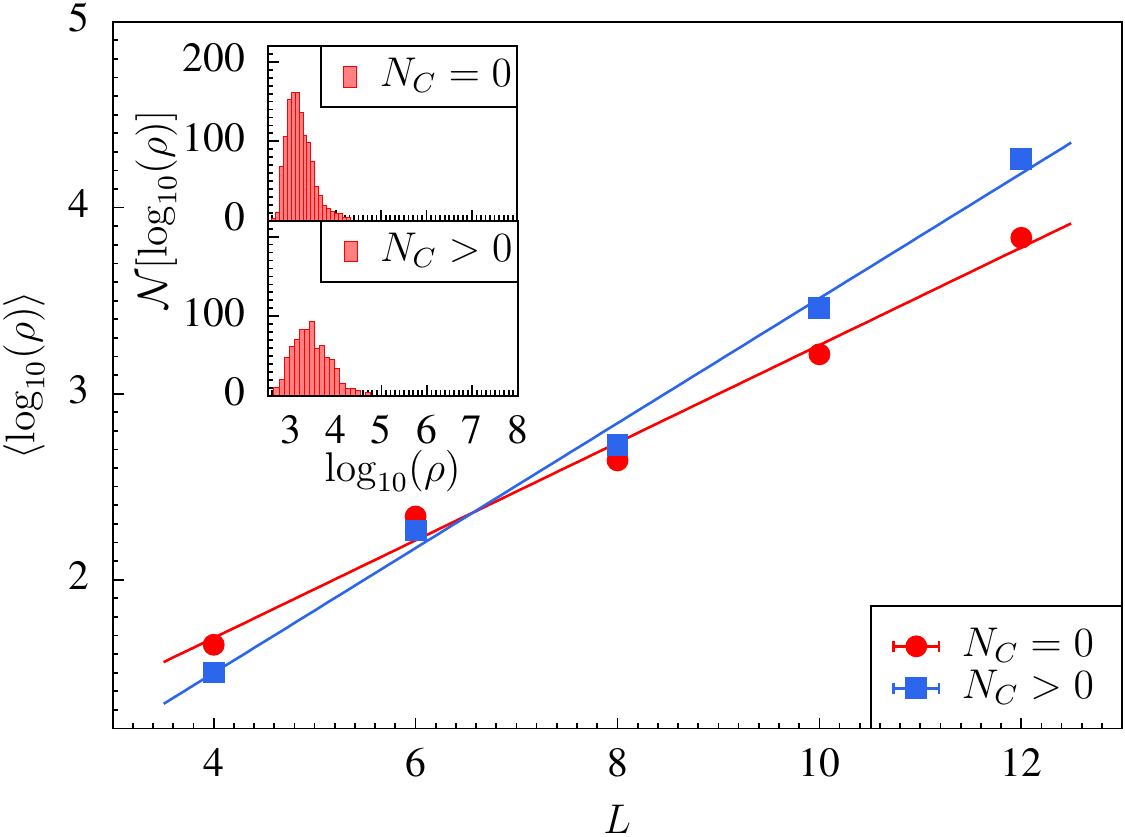}
\caption{(Color online)
The average of the log (base $10$) of the hardness $\rho$, measured at
$\beta=3$, vs size $L$ for two classes of samples, those without
crossing in the range $\beta \in (1.5, 3.0)$, $\nc=0$ (red circles) and
those with at least one crossing in that range, $\nc>0$ (blue squares).
Error bars are smaller than the symbols.  The insets are histograms of
values of $\rho$ for the two classes of samples.
}
\label{fig:rhoL}
\end{center}
\end{figure}

A direct measure of hardness for population annealing for a given sample
is the characteristic family size $\rho$.  In population annealing, most
of the original population is eliminated by successive resampling steps
and the final population is descended from a small subset of the initial
population.  Every member of the final population can be uniquely
assigned to a ``family'' descended from some member of the initial
population. Let $n_i$ be the fraction of the low-temperature population
descended from replica $i$ in the initial population. $\rho$ is then
defined as \begin{equation} \label{ } \rho = \lim_{R_0 \rightarrow
\infty} R_0 \exp\Big[ \sum_i n_i \log n_i\Big], \end{equation} where
$R_0$ is the population size in the simulation ($n_i=0$ there is no
contribution to the sum).  In practice, $\rho$ is measured using the
large $R_0$ of the simulation. Since there may be correlations between
members of the same family, the  population size $R_0$ in the simulation
must be much larger than $\rho$ to assure a large number of independent
measurements and small statistical errors. Thus samples with the largest
$\rho$ require the most computational resources to simulate. We have
shown \cite{wang:14} that $\rho$ is also strongly correlated with the
integrated autocorrelation time for parallel tempering Monte Carlo,
measured in Ref.~\cite{yucesoy:13} so that the same conclusions are
likely to hold for parallel tempering.  Figure \ref{fig:rhoL} shows the
disorder average of $\log_{10} \rho$ vs $L$ measured at $\beta=3$ for
two different classes of disorder samples. The $\nc=0$ class has no
temperature chaos events (crossings with $p_i>0.05$) in the range $\beta
\in (1.5, 3.0)$ while the $\nc>0$ class has one or more temperature
chaos events in the same range.  The error bars are smaller than the
data points and the plots show that $\rho$ scales exponentially in $L$,
$\rho \propto \exp(L/\ell)$, for both classes but that the
characteristic size scale $\ell=1.27(14)$ is significantly smaller for
the chaotic samples than for nonchaotic samples where  $\ell=1.62(10)$.
It is an interesting question whether temperature chaos slows down all
algorithms for spin glasses, not just those that depend on coupling
multiple temperatures.  Recent studies have shown that
\cite{katzgraber:15,hen:15} computationally hard instances for classical
algorithms are also computationally hard for quantum annealing machines,
like the D-Wave Two quantum annealer. As such, by measuring $\rho$ for a
given sample, we have a simple way to uniquely classify the complexity
of a given instance. This means that our approach is of great importance
in the development of hard problems to discern whether quantum annealing
can outperform simulated annealing simulations (see, for example,
Refs.~\cite{nagaj:12,boixo:13a,pudenz:13,boixo:14a,ronnow:14a}).

We have seen that thermal boundary conditions allow us to identify
temperature chaos with boundary condition crossings and provide a tool
for studying chaos quantitatively even for the small sizes accessible to
simulations.   It would be interesting to apply these ideas to other
types of chaos in spin glasses such as bond chaos.  We have also
established that temperature chaos is a significant determinant of
computational hardness for multicanonical algorithms but it remains an
open question as to whether temperature chaos is correlated with
hardness for all algorithms.

\acknowledgements

We thank Alan Middleton and Dan Stein for useful discussions. J.M.~and
W.W.~acknowledge support from National Science Foundation (Grant
No.~DMR-1208046). H.G.K.~also `acknowledges support from the National
Science Foundation (Grant No.~DMR-1151387). H.G.K.'s~research is based
on work supported in part by the Office of the Director of National
Intelligence (ODNI), Intelligence Advanced Research Projects Activity
(IARPA), via MIT Lincoln Laboratory Air Force Contract
No.~FA8721-05-C-0002.  The views and conclusions contained herein are
those of the authors and should not be interpreted as necessarily
representing the official policies or endorsements, either expressed or
implied, of ODNI, IARPA, or the U.S.~Government.  The U.S.~Government is
authorized to reproduce and distribute reprints for Governmental purpose
notwithstanding any copyright annotation thereon.  We thank the Texas
Advanced Computing Center (TACC) at the University of Texas at Austin
for providing HPC resources (Stampede Cluster), and Texas A\&M
University for access to their Ada, Curie, Eos, and Lonestar clusters.

\bibliographystyle{apsrevtitle}
\bibliography{refs}

\end{document}